\documentclass[conference]{IEEEtran}
\usepackage{cite}
\usepackage{amsmath,amssymb,amsfonts}
\usepackage{algorithmic}
\usepackage{graphicx}
\usepackage{textcomp}
\usepackage{xcolor}
\usepackage{textcomp}
\usepackage{url}

\usepackage{censor}
\StopCensoring

\def\BibTeX{{\rm B\kern-.05em{\sc i\kern-.025em b}\kern-.08em
    T\kern-.1667em\lower.7ex\hbox{E}\kern-.125emX}}

\begin{document}

\title{Teaching Code Refactoring using LLMs}
    
\author{
\IEEEauthorblockN{\censor{Anshul Khairnar}}
\IEEEauthorblockA{\censor{\textit{Department of Computer Science}} \\
\censor{\textit{North Carolina State University}}\\
\censor{Raleigh, NC, USA} \\
\censor{akhairn@ncsu.edu}}
\and
\IEEEauthorblockN{\censor{Aarya Rajoju}}
\IEEEauthorblockA{\censor{\textit{Department of Computer Science}} \\
\censor{\textit{North Carolina State University}}\\
\censor{Raleigh, NC, USA} \\
\censor{arajoju@ncsu.edu}}
\and
\IEEEauthorblockN{\censor{Dr. Edward F. Gehringer}}
\IEEEauthorblockA{\censor{\textit{Department of Computer Science}} \\
\censor{\textit{North Carolina State University}}\\
\censor{Raleigh, NC, USA} \\
\censor{efg@ncsu.edu}}
}

\maketitle

\begin{abstract} This Innovative Practice full paper explores how Large Language Models (LLMs) can enhance the teaching of code refactoring in software engineering courses through real-time, context-aware feedback. Refactoring improves code quality but is difficult to teach, especially with complex, real-world codebases. Traditional methods like code reviews and static analysis tools offer limited, inconsistent feedback. Our approach integrates LLM-assisted refactoring into a course project using structured prompts to help students identify and address code smells such as long methods and low cohesion. Implemented in Spring 2025 in a long-lived OSS project, the intervention is evaluated through student feedback and planned analysis of code quality improvements. Findings suggest that LLMs can bridge theoretical and practical learning, supporting a deeper understanding of maintainability and refactoring principles. \end{abstract}


\section{Introduction}
Software maintenance consumes up to 60\% of the total software lifecycle cost, with refactoring being a critical activity for improving code quality, maintainability, and extensibility [1]. Despite the importance of refactoring, teaching effective techniques remains challenging, particularly when students encounter real-world, complex codebases rather than contrived examples [2]. Students often struggle with identifying refactoring opportunities in unfamiliar code and implementing appropriate transformations that preserve functionality while enhancing quality.

Open Source Software (OSS) projects offer an authentic environment for students to practice refactoring skills. However, the complexity of these projects can overwhelm students, and providing individualized feedback at scale becomes a significant challenge for instructors [3]. Traditional approaches rely heavily on code reviews, static analysis tools, and manual feedback, which can be time-consuming and inconsistent [4].

The emergence of Large Language Models (LLMs) presents a promising opportunity in software engineering education. These models demonstrate remarkable capabilities in understanding, generating, and transforming code [5]. Unlike traditional static analysis tools that rely on predefined patterns, LLMs can reason about code semantics, understand complex contexts, and provide detailed explanations—capabilities that align well with the needs of refactoring education.

This paper presents an innovative approach to teaching refactoring using LLMs applied to OSS projects. Our work is guided by the following research questions:

\begin{enumerate}
\item How effective are LLMs in identifying opportunities for refactoring compared to human developers?
\item What types of refactoring tasks (renaming variables, extracting methods, etc.) do LLMs perform well, and which ones do they struggle with?
\item Having used an LLM to assist in refactoring, how likely is a programmer to continue to use LLMs for that task?
\end{enumerate}

Our approach involves applying LLM-assisted refactoring to \censor{Expertiza}, a widely used and actively maintained OSS application which is used to facilitate collaborative learning and peer assessment in an Object-Oriented Design and Development course. The study was implemented in the Spring 2025 offering of the course, allowing us to gather comprehensive data on the effectiveness of this approach. The students undertaking this particular course work on this open-source project as part of their coursework, and anyone doing a refactoring project is required to use LLMs. We will be surveying these participants for their feedback.

This paper contributes to the field by: (1) designing and implementing an LLM-assisted approach for teaching code refactoring in OSS projects, (2) evaluating the effectiveness of LLMs in identifying and implementing various types of refactorings, (3) assessing student experiences and perceptions regarding the use of LLMs for refactoring tasks, and (4) providing insights and recommendations for educators interested in incorporating LLMs into software engineering education.

\section{Background and Literature Review}

\subsection{Refactoring: Principles and Challenges}

Refactoring, as defined by Fowler et al. [6], is the process of changing a software system to improve its internal structure without altering its external behavior. The primary goal is to enhance code quality attributes such as readability, maintainability, and extensibility while preserving functionality. Common refactoring techniques include extracting methods, renaming variables, moving methods between classes, and simplifying conditional expressions [7].

Despite its importance, effective refactoring remains challenging, particularly for novice developers. Murphy-Hill et al. [4] found that developers often struggle with identifying refactoring opportunities, selecting appropriate techniques, and verifying correctness. These challenges are amplified when dealing with large, complex codebases typical of real-world software projects [8]. Additionally, traditional software engineering education often emphasizes new feature development over maintenance and refactoring [9], resulting in graduates who are ill-prepared for the realities of software maintenance.

\subsection{Teaching Refactoring: Traditional Approaches and Limitations}

Traditional approaches to teaching refactoring include classroom instruction, code-review exercises, and refactoring assignments using small, contrived examples [10]. While these methods provide foundational knowledge, they often fail to capture the complexity and challenges of refactoring in real-world contexts [11].

Several educational initiatives have attempted to address these limitations by incorporating OSS projects into refactoring education. For instance, Szabo et al. [12] proposed a pedagogical approach where students identify and refactor code smells in OSS codebases. Similarly, Pinto et al. [13] developed a framework for teaching refactoring through collaborative OSS contributions. These approaches provide students with authentic experiences but typically rely on manual guidance and feedback from instructors, which can be time-consuming and inconsistent.

Automated tools such as static analyzers and refactoring assistants have been employed to support refactoring education. Tools like JDeodorant [14] and SonarQube help detect code smells and suggest refactoring opportunities. However, these tools often generate false positives, require significant configuration, and may not adequately explain the rationale behind refactoring suggestions [15].

\subsection{LLMs in Software Engineering Education}

The integration of LLMs into software engineering education represents an emerging trend with significant potential. Recent studies have explored the use of LLMs for tasks such as code generation [16], debugging [17], and program understanding [18]. These models demonstrate capabilities that could transform how programming and software engineering are taught and learned.

In the educational context, LLMs offer several advantages: they can provide personalized feedback at scale, generate examples tailored to specific learning objectives, and simulate expert reasoning [19]. However, concerns persist regarding accuracy, reliability, and the potential to reinforce misconceptions [20]. Additionally, there is an ongoing debate about how to effectively integrate LLMs into curriculum design to enhance rather than replace critical thinking and problem-solving skills [21].

\subsection{LLMs for Code Refactoring}

Recent research has begun to explore the potential of LLMs for code refactoring tasks. Zhang et al. [22] developed a model that uses LLM-generated information to recommend move method refactoring with improved precision and recall compared to traditional methods. Similarly, Zhang et al. [23] proposed a hybrid knowledge-driven approach leveraging LLMs for refactoring to Pythonic idioms, achieving high accuracy and maintainability improvement.

LLMs offer several advantages for refactoring tasks. They can understand code semantics beyond syntactic patterns, learn from diverse examples across different programming languages and domains, and generate detailed explanations for their suggestions \cite{24,25,26,27,28}. These capabilities make them potential tools for both identifying refactoring opportunities and implementing refactorings.

However, challenges remain. Recent empirical studies have revealed that LLMs may struggle with complex architectural refactorings, ensuring behavior preservation, and understanding project-specific contexts. For instance, a study by Warden found that LLM-generated refactorings were correct only in 37\% of cases without additional fact-checking mechanisms. With fact-checking incorporated, the correctness rate increased to 98\%, highlighting the importance of verification in LLM-assisted refactoring.

\subsection{Bridging the Gap: LLMs for Teaching Refactoring}

Despite the growing body of literature on both teaching refactoring and using LLMs for refactoring tasks, little research has explicitly explored the intersection of these domains. The use of LLMs as pedagogical tools for teaching refactoring represents a promising yet underexplored area.

A few recent studies have begun to bridge this gap. For example, Wu et al. explored the use of LLMs to generate refactoring examples for educational purposes, while Ren et al. investigated how LLM-generated explanations could enhance student understanding of refactoring principles. However, comprehensive approaches that leverage LLMs to support the entire refactoring learning process—from identifying opportunities to implementing and verifying refactorings in OSS contexts—remain largely unexplored.

This gap is particularly notable given the potential synergies between LLMs' capabilities and the challenges of teaching refactoring. LLMs could provide personalized, context-aware guidance that addresses the specific difficulties students face when learning to refactor complex codebases. By combining the authenticity of OSS projects with the adaptive support of LLMs, educators could potentially offer more effective and scalable refactoring education.

Our work aims to address this gap by designing, implementing, and evaluating an LLM-assisted approach to teaching refactoring in OSS contexts. By systematically investigating the effectiveness of LLMs for different refactoring tasks and assessing student experiences, we contribute to both the theoretical understanding and practical applications of LLMs in software engineering education.

\section{Methodology}

This research is conducted as part of a graduate software engineering course, where students conduct refactoring of a real Open-Source Software (OSS) project with the assistance of Large Language Models (LLMs). The objective is to examine how LLMs assist in improving the readability, maintainability, and adherence to best practices of the code.

\subsection{Project Setup}
\censor{Expertiza} was selected as the medium-complexity OSS project, due to the instructor's long history with it. Students taking the course \censor{CSC 517 (Object-Oriented Design and Development)} at \censor{North Carolina State} University are tasked with contributing to this OSS project as part of their coursework. The students work in teams and choose various code modules from this codebase to refactor.

\subsection{LLM Integration}
Students are required to use an LLM in refactoring.  They are encouraged to use tools such as ChatGPT and GitHub Copilot, with prepared prompt templates that are designed to tackle common refactoring needs, such as long methods, deeply nested conditionals, duplicated code, and low cohesion. Students were provided with a 3-page document of suggestions on how to write prompts (see Appendix~\ref{appendix:tips} for a shortened version).

\subsection{Refactoring Workflow}
The refactoring process consists of the following steps:
\begin{enumerate}
\item Identify problematic code through manual inspection and static analysis.
\item Apply LLM-generated suggestions to increase structure and ease of maintenance.
\item Record all refactoring, including the prompt used, output, and manual revisions applied.
\item Commit changes in code through version control with a reflection log containing the rationale and result.
\end{enumerate}

\subsection{Survey Design and IRB Approval}

To evaluate the impact of LLM-assisted refactoring, we administered two surveys: a pre-survey before the project began and a post-survey upon its completion. The pre-survey focused on prior knowledge of refactoring concepts, comfort with legacy code, and LLM experience. The post-survey captured reflections on LLM usage, benefits, limitations, and learning outcomes.

This study was reviewed and approved by the \censor{North Carolina State} University Institutional Review Board. Participation was voluntary and anonymous.

\subsection{Data Collection}

Student feedback data was collected via a survey using Google Forms and included both multiple-choice and open-ended questions. Student reflection logs and selected refactored code submissions were also collected for exploratory qualitative analysis.

\subsection{Instructor Feedback}
Refactored code projects were assessed by instructors and teaching assistants. These reviews were used to evaluate the correctness of the refactorings, adherence to design principles such as modularity and cohesion, and the reasoning students applied in making structural improvements. The takeaways from the instructor are summarized in section~\ref{discussion:instructor_takeaways}.

\section{Evaluation and Results}

The evaluation of LLM-facilitated refactoring was mainly qualitative and was collected through three major sources: pre-surveys, post-surveys, and student reflection logs. These sources were used to examine the support provided by LLMs to students in understanding, planning, and executing code refactorings in a large, open-source environment. The open-ended responses and student reflections were analyzed using informal qualitative methods. Two members of the research team independently reviewed responses and discussed recurring patterns, but no formal coding rubric was used. As such, we describe our analysis as exploratory and interpretive.

\subsection{Survey}

After completing the project, students completed a post-survey to capture their opinions on LLM usage, perceived benefits, challenges faced, and overall learning experience. The post-survey included both multiple-choice (quantitative) and open-ended (qualitative) questions.

These key areas were addressed:
\begin{itemize}
    \item Challenges of refactoring faced in large, interdependent codebases.
    \item Application of LLMs and contexts where they were most beneficial.
    \item Perceived effectiveness of LLMs in improving code readability, structure, and maintainability.
    \item Suggestions for improving the utility and usability of LLM tools in software development activities.
\end{itemize}

\subsection{Response Rate}
Out of the 46 enrolled students taking part in refactoring projects, 24 submitted valid responses to the post-survey, yielding a response rate of approximately 52\%.

\subsection{Observations from the survey results}

\begin{enumerate}
    \item Among the students surveyed, a majority (70\%) reported using OpenAI’s GPT models for their refactoring tasks, with Gemini, and Claude also being commonly used alternatives as illustrated in Figure~\ref{fig:llms_used}.

\begin{figure}[h]
    \centering
    \includegraphics[width=0.5\textwidth]{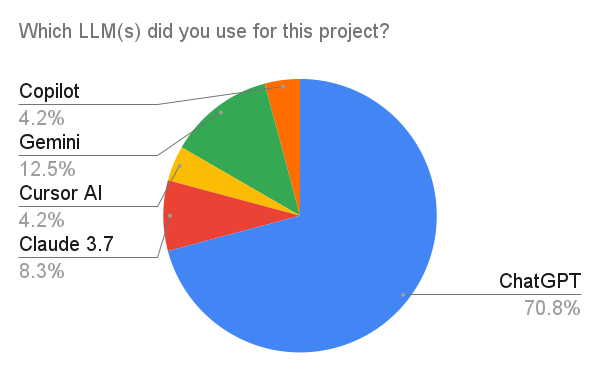}
    \caption{Various LLMs used by the students (\textit{n=24})}
    \label{fig:llms_used}
\end{figure}

    \item Usage frequency varied, but the majority of students interacted with LLMs occasionally throughout the project. A notable subset used them frequently, while only a few relied on them rarely (see Figure~\ref{fig:freq}), indicating that most students treated LLMs as supportive tools rather than primary drivers in their refactoring process.

\begin{figure}[h]
    \centering
    \includegraphics[width=0.5\textwidth]{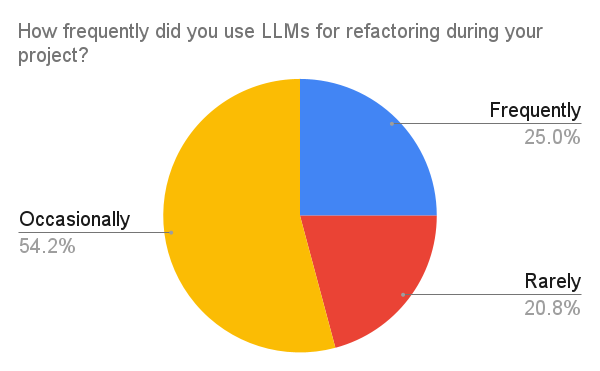}
    \caption{Frequency of LLM usage (\textit{n=24})}
    \label{fig:freq}
\end{figure}

    \item As shown in Figure~\ref{fig:behaviour}, most students encountered occasional errors or unexpected behaviors in LLM-generated refactorings. A few students reported frequent problems, while some experienced no difficulties at all or were unsure. This variation suggests that while LLMs were generally helpful, their outputs still required careful review and testing to ensure correctness..

\begin{figure}[h]
    \centering
    \includegraphics[width=0.5\textwidth]{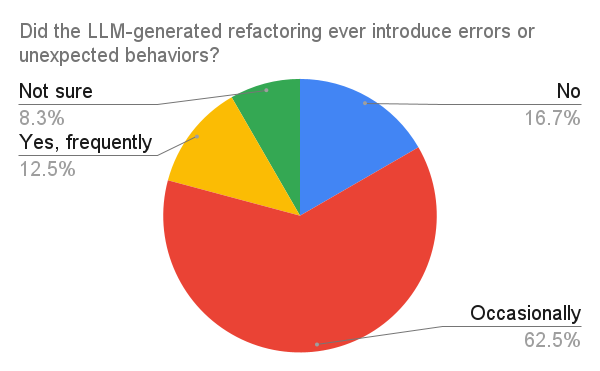}
    \caption{Errors or Unexpected behaviors (\textit{n=24})}
    \label{fig:behaviour}
\end{figure}

    \item Figure~\ref{fig:helpful} shows that students generally found LLM-based refactoring to be moderately to highly helpful for their projects. Most ratings clustered around 3 and 4, with a few students rating the experience as extremely helpful (5) and only one rating it lower (2). This suggests that while the support from LLMs was not flawless, it was widely perceived as a valuable aid in the refactoring process.

\begin{figure}[h]
    \centering
    \includegraphics[width=0.5\textwidth]{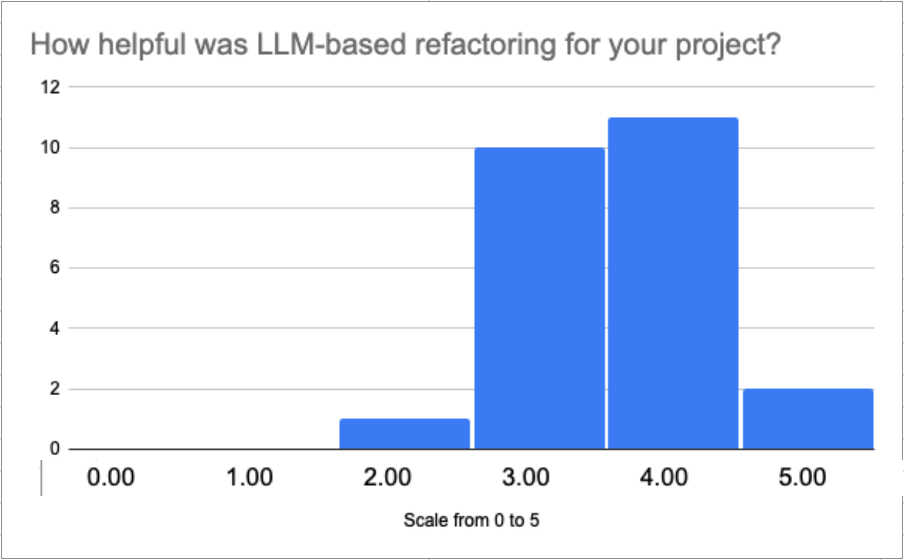}
    \caption{Helpfulness of LLMs in refactoring (\textit{n=24})}
    \label{fig:helpful}
\end{figure}

    \item While many students reported that the LLM provided accurate refactoring suggestions within 1–3 attempts, success on the first try was often dependent on the complexity of the code and the clarity of the prompt. As reflected in Figure~\ref{fig:attempts}, students frequently had to adjust their prompts—by adding contextual details, clarifying expectations, or pointing out specific issues—to guide the LLM toward a correct and useful output. This highlights the importance of prompt engineering as a critical skill in effectively leveraging LLMs for code refactoring.

\begin{figure}[h]
    \centering
    \includegraphics[width=0.5\textwidth]{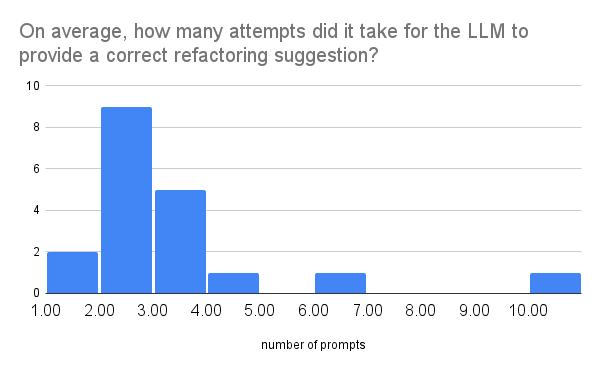}
    \caption{Number of attempts for a correct suggestion (\textit{n=24})}
    \label{fig:attempts}
\end{figure}

    \item As illustrated in Figure~\ref{fig:future}, the majority of students expressed a willingness to continue using LLMs for refactoring in future projects. While several students selected “Maybe,” indicating conditional interest based on improvements or context, only a small number responded with a definitive “No.” This suggests that overall sentiment toward LLM-assisted refactoring remains positive, with most students seeing potential for future integration.

\begin{figure}[h]
    \centering
    \includegraphics[width=0.5\textwidth]{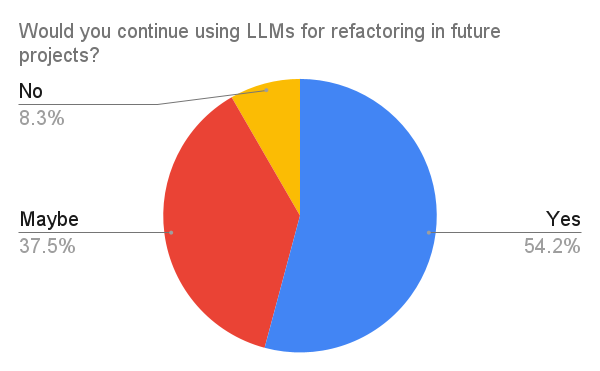}
    \caption{Continued usage of LLMs in the future (\textit{n=24})}
    \label{fig:future}
\end{figure}

\end{enumerate}

\subsection{Results}

Comparison of the survey results led to several important observations derived from open-ended survey responses and student reflection logs, which were reviewed manually for recurring ideas. 

\begin{itemize}
    \item \textit{Improved refactoring with LLMs:} Students reported that LLMs were helpful in deconstructing code with numerous or unfamiliar components, identifying repetitive code, and offering improved naming or structural suggestions.
    
    \item \textit{Increased efficiency with human input:} While LLMs reduced time spent on understanding and planning in many cases, students noted that LLM-generated code often required manual modifications, especially in cases involving project-specific rules or domain-specific logic.
    
    \item \textit{Prompt engineering became a skill:} Most students found that refining their prompts significantly improved the usefulness of LLM responses. They used multiple iterations to direct the LLM toward more precise and context-aware outputs.
    
    \item \textit{Critical thinking in code reviews:} Students did not accept LLM suggestions blindly. They made informed overrides, merged ideas from different outputs, or used LLMs as thought partners rather than final decision-makers.
    
    \item \textit{Aspiration for tooling improvement:} A recurring theme in feedback was the desire for more project-aware suggestions and better integration with development environments such as GitHub and VS Code.
    
    \item \textit{Positive outlook for future use:} Despite the drawbacks, the majority of students expressed willingness to use LLMs in future refactoring projects, particularly as aids in understanding large codebases and identifying areas for structural improvement.
\end{itemize}

\subsection{Research Questions}

\begin{itemize}
    \item \textit{RQ1: How effective are LLMs in identifying refactoring opportunities compared to human developers?} Most students saw LLMs as a fast and helpful starting point. The models were great at spotting low-hanging fruit but still relied on human review to catch edge cases or ensure that suggested changes aligned with the overall project structure.

    \item \textit{RQ2: Which refactoring tasks do LLMs handle well or struggle with?} Students reported that LLMs did a solid job at localized tasks---things like variable renaming, extracting helper functions, and cleaning up indentation or formatting. However, the tools struggled when changes involved larger architectural decisions, like breaking up monolithic classes or preserving domain-specific conventions.  

    \item \textit{RQ3: Would students use LLMs for refactoring in the future?}  The response was largely positive. Most students said they’d definitely use an LLM again, especially for brainstorming or early-stage cleanup. However, they emphasized the importance of using them alongside testing tools and maintaining a healthy level of skepticism.
\end{itemize}

\section{Discussion}

\subsection{Findings}

The findings of this study show the potential of LLMs as supportive instruments for teaching code refactoring. While students benefited from faster comprehension and improved initial structuring, the greatest gains in learning were due to active reflection on the tools by the students.

Students encountered a variety of challenges during refactoring, with the most common being understanding and navigating a large, unfamiliar codebase—especially one written in Ruby on Rails, which many had not worked with before. Several students noted that working with legacy code, resolving interdependencies, and ensuring their changes didn’t break other parts of the system were particularly difficult. On the LLM side, common issues included hallucinated suggestions, generic or incorrect refactorings, and the need for prompt refinement to achieve accurate results. These challenges highlight the dual complexity of refactoring real-world systems and effectively collaborating with AI tools in practice.

Here are some of the common ideas we observed:

\subsubsection{Human-AI collaboration}
One of the prominent themes to emerge from our surveys and look-back was the interaction between human judgment and automation. Students would usually use LLMs to initiate refactoring tasks—particularly detecting similar code or naming conventions—but left it up to their own judgment to apply domain-specific constraints. This aligns with previous work that supports human-in-the-loop AI in software development.

\subsubsection{Prompt engineering as a learning outcome}
The students who successively refined prompts and included context-related instructions reported higher satisfaction and lower error rates. This supports the idea that prompt engineering should be seen as a key skill in modern software engineering education.

\subsubsection{Mixed effectiveness and critical reflection}
A large majority of students felt that the refactoring process was faster and more efficient with LLMs, and many also noted that it led to higher-quality code. Several responses highlighted that LLMs helped them better understand the purpose and flow of existing code, which accelerated decision-making. However, a few students also observed that LLMs occasionally introduced unnecessary complexity or had no significant impact on quality. These mixed experiences underscore that while LLMs can be powerful aids, their effectiveness depends on thoughtful use and careful review.

\subsection{Implication for Tool Design}
When asked about potential improvements to LLM-based refactoring tools, students overwhelmingly highlighted the need for more context-aware suggestions that consider project structure. Many also emphasized the value of seamless integration with development workflows (e.g., GitHub, VS Code), better explanations for suggested changes, and customizable prompting features. This indicates a strong demand for LLM tools that go beyond isolated code snippets and offer deeper, project-level understanding and developer-centric usability.

\subsection{Implications for Software Engineering Education}
Bringing LLMs into the classroom opens up exciting opportunities. Students receive immediate feedback, learn from examples, and explore alternative approaches. For instructors, it offers a scalable way to help students practice applying design principles in real code. However, LLMs are not a plug-and-play solution; effective integration requires deliberate instructional design and scaffolding. Students need clear expectations, reflective prompts, and a safe space to experiment, fail, and learn.

\subsection{Implication for Assessment}
As AI becomes embedded in programming practice, assessment models must evolve. Rather than evaluating only final outputs, instructors may need to assess students’ ability to critically engage with AI-generated suggestions, justify design decisions, and demonstrate an understanding of trade-offs.

\subsection{Balancing Automation and Learning}
An important concern in AI-supported education is the risk of students completely outsourcing their thinking to these AI tools. Our design countered this by encouraging students not to accept the first AI response and by requiring peer reviews of designs as part of the course. These practices appear to have effectively encouraged metacognitive reflection. Students not only used LLMs to generate refactorings but also learned to critique them. This balance enabled increased productivity without sacrificing learning outcomes.

\subsection{Instructor's Takeaways}  \label{discussion:instructor_takeaways}
LLMs proved to be very effective for within-class refactorings. They encouraged good naming practices and breaking up of complex methods.  They were much less effective at class-to-class refactorings.  They almost never suggested moving functionality from its current class to another class where it would increase cohesion or diminish coupling.  Unfortunately, for the Expertiza project at least, the main reason the code is hard to understand is that related responsibilities are distributed around the system instead of concentrated in a single class where they would be easier to comprehend. Nonetheless, the refactorings suggested by these tools would enhance the clarity of large parts of the code, making it easier to see where specific functionalities could be moved.

These results suggest that LLMs, when used reflectively and critically, can serve as effective scaffolding tools to support student learning of software maintainability and refactoring in real-world settings.

\subsection{Threats to Validity}
This was a single-course deployment with one cohort, so our findings may not generalize to all learning environments. With only 24 responses, the sample size is relatively small. Additionally, because we focused mostly on qualitative data (surveys, reflections, instructor reviews), we do not have concrete measurements of how much code quality improved. Some teams used the LLM extensively; others barely touched it. This variation will be addressed in future iterations through more structured check-ins and logging. 

\section{Conclusion}
Teaching students to refactor large, messy codebases is tough, but LLMs offer a new way to make it easier, faster, and more engaging. In this study, we saw that when used thoughtfully, LLMs helped students identify and carry out meaningful code improvements. More importantly, students learned to treat AI tools as a collaborator, and not just a shortcut, and they also developed stronger judgment about what good refactoring looks like.

When students reason as to why they applied (or ignored) a suggestion, they build lasting understanding. With the right balance of automation and reflection, LLMs can become not just a coding aid, but a valuable teaching partner.

As LLMs become more integrated into professional software workflows, helping students learn how to collaborate with these tools, rather than depend on them, is becoming an essential part of software engineering education.

\section{Future Work}

We plan to complement our qualitative findings from this study with quantitative code quality metrics. These will include measures such as cyclomatic complexity, code duplication, coupling and cohesion, and technical debt reduction metrics. We also plan to conduct a deeper analysis of how student teams iteratively refined LLM prompts and how those refinements correlated with refactoring quality.

In the long term, we aim to expand this work across multiple course offerings. We also hope to explore IDE-integrated LLM tooling that provides just-in-time suggestions during development and to measure how student trust and reliance on LLMs evolve over time.

\newpage

\section*{Acknowledgments}

We are grateful to all participants and collaborators who contributed to this study. Their insights and engagement were invaluable in shaping the research direction and outcomes.
We also acknowledge the use of AI tools, including OpenAI's GPT models, for assisting in this study, by analyzing literature review, refining prompts, generating textual information, and analyzing refactoring scenarios in this study.

\newpage

\appendices
\section{\textbf{Tips for Using LLMs to Refactor Code}} \label{appendix:tips}

\subsection{General strategy}

When refactoring code, the goal is to improve readability, maintainability, and efficiency while preserving functionality. Large Language Models (LLMs) can assist in the following ways:

\begin{itemize}
    \item \textbf{Understanding code} – Break down what each function does.
    \item \textbf{Identifying code smells} – Detect duplicated logic, long functions, etc.
    \item \textbf{Suggesting improvements} – Rewrite code using best practices.
    \item \textbf{Step-by-Step refactoring} – Modify incrementally to avoid breaking functionality.
    \item \textbf{Validating refactored code} – Ensure correctness and minimal performance impact.
\end{itemize}

\subsection{General prompt examples for refactoring}

\subsubsection{Understanding the existing code}

\begin{itemize}
    \item ``Can you explain what this class does step by step?''
    \item ``Identify all dependencies in this class and how they interact.''
    \item ``Summarize the core logic of this class in simpler terms.''
\end{itemize}

\subsubsection{Detecting code smells}

\begin{itemize}
    \item ``What potential code smells do you see in this class?''
    \item ``Does this class follow the Single Responsibility Principle?''
    \item ``Are there any redundant database queries or performance bottlenecks?''
\end{itemize}

\subsubsection{Suggesting improvements}

\begin{itemize}
    \item ``Refactor this class to improve readability and maintainability.''
    \item ``Can you split this class into smaller sub-classes without changing behavior?''
    \item ``Rewrite this code using design patterns like the 23 Gang of Four patterns.''
\end{itemize}

\subsubsection{Step-by-step refactoring}

\begin{itemize}
    \item ``Refactor this function gradually, explaining each step and ensuring correctness.''
    \item ``Rewrite this method to be more modular while keeping the same input-output behavior.''
    \item ``Can you reduce code duplication in this file while keeping the logic intact?''
\end{itemize}

\subsubsection{Validating and testing refactored code}

\begin{itemize}
    \item ``Provide test cases to verify the correctness of the refactored class.''
    \item ``Check if the new implementation maintains the same functionality as the original.''
    \item ``Are there any edge cases that this class does not handle?''
\end{itemize}

\subsection{Specific refactoring techniques evaluated through literature review}

\subsubsection{One-Shot Prompting}
Provide a single example of a refactored method and ask the LLM to refactor a similar method:

\begin{quote}
``Here's an example of a refactored method: [Example]. Now, refactor this method in a similar way: [Method to refactor]''
\end{quote}

\subsubsection{Few-Shot Prompting}
Supply multiple examples of refactored code to guide the LLM:

\begin{quote}
``Here are three examples of refactored methods: [Examples]. Using these as guidelines, refactor the following method: [Method to refactor]''
\end{quote}

\subsubsection{Chain-of-Thought Prompting}
Guide the LLM through a step-by-step refactoring process:

\begin{quote}
``Let's refactor this method step by step:
\begin{enumerate}
    \item Identify any code smells in the method.
    \item Suggest how to address each code smell.
    \item Rewrite the method, explaining each change.
    \item Verify that the new method maintains the original functionality.
\end{enumerate}
Now, start with step 1 for this method: [Method to refactor]''
\end{quote}

\subsection{Additional examples of prompting}

\subsubsection{Incremental refactoring}
\begin{itemize}
    \item ``Refactor only one class/method at a time and explain why the change is needed.''
    \item ``List the top three improvements for this file and implement them step by step.''
\end{itemize}

\subsubsection{Context-aware refactoring}
\begin{itemize}
    \item ``How does this class fit into the larger project? Suggest improvements in that context.''
    \item ``Given that this project uses [Rails/Node.js], how can this class follow best practices?''
\end{itemize}

\subsubsection{Style guide compliance}
\begin{itemize}
    \item ``Modify this code to follow Ruby on Rails best practices.''
    \item ``Ensure this class adheres to SOLID principles.''
\end{itemize}

\end{document}